\documentstyle[12pt,epsf]{article} 
\catcode`@=11
\newcount\@tempcntc
\def\@citex[#1]#2{\if@filesw\immediate\write\@auxout{\string\citation{#2}}\fi
  \@tempcnta\z@\@tempcntb\m@ne\def\@citea{}\@cite{\@for\@citeb:=#2\do
    {\@ifundefined
       {b@\@citeb}{\@citeo\@tempcntb\m@ne\@citea\def\@citea{,}{\bf ?}\@warning
       {Citation `\@citeb' on page \thepage \space undefined}}%
    {\setbox\z@\hbox{\global\@tempcntc0\csname b@\@citeb\endcsname\relax}%
     \ifnum\@tempcntc=\z@ \@citeo\@tempcntb\m@ne
       \@citea\def\@citea{,}\hbox{\csname b@\@citeb\endcsname}%
     \else
      \advance\@tempcntb\@ne
      \ifnum\@tempcntb=\@tempcntc
      \else\advance\@tempcntb\m@ne\@citeo
      \@tempcnta\@tempcntc\@tempcntb\@tempcntc\fi\fi}}\@citeo}{#1}}
\def\@citeo{\ifnum\@tempcnta>\@tempcntb\else\@citea\def\@citea{,}%
  \ifnum\@tempcnta=\@tempcntb\the\@tempcnta\else
   {\advance\@tempcnta\@ne\ifnum\@tempcnta=\@tempcntb \else \def\@citea{--}\fi
    \advance\@tempcnta\m@ne\the\@tempcnta\@citea\the\@tempcntb}\fi\fi}
\catcode`@=12

\begin{document}
\vskip -1.5cm
\begin{flushright}
WUE-ITP-2001-004\\[-0.1cm] 
hep-ph/0101266\\[-0.1cm]
January 2001
\end{flushright}

\begin{center}
{\Large {\bf Light Charged Higgs Boson and Supersymmetry}}\\[1.7cm]
{\large C. Panagiotakopoulos$^a$ and A. Pilaftsis$^b$}\\[0.4cm]
$^a${\em Physics Division, School of Technology,
         Aristotle University of Thessaloniki,\\
         54006 Thessaloniki, Greece}\\[0.2cm]
$^b${\em Institut f\"ur Theoretische Physik und Astrophysik,
         Universit\"at W\"urzburg,\\
         Am Hubland, 97074 W\"urzburg, Germany}    
\end{center}
\vskip1.cm  \centerline{\bf ABSTRACT}  
A possible  discovery of a  relatively light charged Higgs boson $H^+$
in near future experiments, with a mass $M_{H^+} \stackrel{<}{{}_\sim}
110$~GeV, together with the present LEP2 direct limits on the chargino
and neutral Higgs sectors,  would disfavour the minimal supersymmetric
standard  model  as well  as  its frequently discussed next-to-minimal
supersymmetric extension.   We show that   a supersymmetric origin can
naturally be ascribed  to the existence of  such a light charged Higgs
scalar  within the   context   of  the  recently   introduced  minimal
nonminimal supersymmetric standard model.

\newpage

Supersymmetry (SUSY) appears to  be a compelling ingredient  of string
theories which are expected  to successfully describe the Planck-scale
dynamics, thereby aspiring to unify  all fundamental forces in nature,
including gravity. For these reasons,  low-energy realizations of SUSY
softly  broken   at   0.1--1   TeV energies,   such  as  the   Minimal
Supersymmetric Standard Model (MSSM)  and its minimal  extensions, are
considered  to  be the   best-motivated models~\cite{SUSY} of  physics
beyond the  Standard Model (SM).   Most interestingly, such low-energy
realizations of SUSY exhibit gauge-coupling unification~\cite{CPW} and
can solve, at least technically, the problem of perturbative stability
of radiative effects between   the  soft SUSY-breaking scale   $M_{\rm
  SUSY}  \sim 1$~TeV   and   the Planck  mass   $M_{\rm  P}$.  These
appealing properties of low-energy SUSY  might be considered to mainly
emanate  from the  doubling of  the particle  spectrum of the  SM; the
theory introduces  a new fermion (boson)  for each SM boson (fermion),
its so-called superpartner.  Superpartners  have typical masses of the
order of the soft  SUSY-breaking scale $M_{\rm   SUSY}$ and should  be
heavier than $\sim 100$~GeV,  for phenomenological reasons~\cite{PDG}. 
In addition, within  the framework of SUSY,  the holomorphicity of the
superpotential together with   the requirement of cancellation  of the
triangle gauge  anomalies entail that the SM  Higgs sector itself must
be augmented  by at least one  Higgs doublet of  opposite hypercharge. 
To  be specific, low-energy SUSY models  include  a minimal set of two
Higgs   iso-doublets and so  necessarily   predict the existence of at
least one  (doublet) charged Higgs  boson, $H^\pm$,  in addition  to a
number of neutral  Higgs particles.  As we  will see in this note, the
mass of $H^+$ introduces a new scale into the neutral Higgs-boson mass
spectrum  and can play a  key r\^ole in distinguishing among different
minimal models of electroweak-scale SUSY.

In  the decoupling limit of  a  heavy charged Higgs boson,  low-energy
SUSY models make  a definite prediction for  the mass of  that neutral
Higgs boson    $H$,  which  is  predominantly  responsible    for  the
spontaneous breaking of the SM gauge group.  In other words, for large
values  of $M_{H^+}$,  e.g.\  $M_{H^+} \sim   1$~TeV, the mass  of the
SM-like   $H$   boson    reaches     a  calculable     model-dependent
maximum~\cite{EQ}.  For instance, in the  MSSM, with radiative effects
included \cite{SHiggs}, recent computations~\cite{CQW,HHW,CHHHWW} lead
to   the  upper limit:  $M_H  \stackrel{<}{{}_\sim} 110$~(130)~GeV for
$\tan\beta \approx  2\ (\stackrel{>}{{}_\sim}  10)$, where $\tan\beta$
is the ratio of the vacuum expectation values (VEV's) of the two Higgs
doublets.  On the other hand, in the frequently discussed extension of
the MSSM known   as the Next-to-Minimal Supersymmetric Standard  Model
(NMSSM) \cite{nmssm}, the maximum  of the corresponding $H$-boson mass
may increase by an amount  of $\sim 30$~GeV, for $\tan\beta\approx 2$,
while     it       remains    unaffected for    large       values  of
$\tan\beta$~\cite{EKW}.       Apart   from $\tan\beta$,   however, the
Higgs-boson mass spectrum depends very sensitively on the actual value
of    $M_{H^+}$   and the   stop-mixing   parameter    $X_t  = A_t   -
\mu/\tan\beta$, where $A_t$ is  the soft SUSY-breaking Yukawa coupling
to stops  and $\mu$  the mixing  parameter  of the  two  Higgs-doublet
superfields  $\widehat{H}_1$  and  $\widehat{H}_2$.  For  example, for
$M_{H^+}  \stackrel{<}{{}_\sim} 110$~GeV, the  masses  of the lightest
CP-even Higgs boson~$h$ and the CP-odd scalar~$A$  are predicted to be
both less  than $\sim  80$~GeV,  almost independently  of $\tan\beta$,
provided no unusually large  values  of $|\mu|$ are considered,  e.g.\
for $|\mu|  \stackrel{<}{{}_\sim} 2\,M_{\rm  SUSY}$.  However,  such a
scenario is disfavoured by the  latest LEP2 data~\cite{ADLO}, since it
predicts an enhanced  $ZhA$-coupling,  whenever the $hZZ$-coupling  is
suppressed, and  hence would have been  detected in  the corresponding
$ZhA$  channel.  As has been  explicitly  demonstrated in \cite{PP}, a
similar negative conclusion may be  reached in the  NMSSM as well, for
$M_{H^+}\stackrel{<}{{}_\sim}  110$~GeV.  In  this model,  the SM-like
Higgs boson $H$  always comes out to  be lighter  than $H^+$, provided
the    effectively     generated    $\mu$-parameter lies     in    the
phenomenologically   favoured  range,     $|\mu| \stackrel{>}{{}_\sim}
100$~GeV,   as  is   suggested  by  the  non-observation  of  chargino
production at LEP2~\cite{PDG,GKLR}.\footnote[1]{Throughout the paper,
  we  shall   not   consider  possible   indirect constraints  on  the
  $H^+$-boson mass from $b\to s\gamma$ and other observables involving
  $B$  mesons, as the  derived  limits sensitively  depend  on several
  other model-dependent parameters of the theory, such  as the sign of
  $\mu$~\cite{BBMR} and the low-energy flavour-mixing structure of the
  squark sector~\cite{NS}.}

Given the difficulty that the MSSM and NMSSM cannot easily accommodate
a charged Higgs boson $H^+$   lighter than the SM-like neutral   Higgs
boson $H$, one may  now raise the  following  question: Should  such a
light charged  particle, with $M_{H^+} \stackrel{<}{{}_\sim} 110$~GeV,
be observed,  e.g.\ at  the  upgraded  Tevatron collider,  is it  then
possible to  ascribe  to it  a supersymmetric  origin within a minimal
SUSY  extension of the  SM?   In this note  we  address this important
question in  the  affirmative  within  the framework of  the  recently
introduced    Minimal  Nonminimal   Supersymmetric   Standard    Model
(MNSSM)~\cite{PP,PT}.

In   the MNSSM the  $\mu$-parameter is   promoted  to a chiral singlet
superfield   $\widehat{S}$, and   all   linear,  quadratic  and  cubic
operators   involving  only     $\widehat{S}$  are    absent  from the
renormalizable superpotential; $\widehat{S}$ enters through the single
term   $\lambda\,  \widehat{S}\,\widehat{H}_1 \widehat{H}_2$.      The
crucial difference between  the MNSSM and the  NMSSM lies in the  fact
that the cubic term    $\frac{1}{3}\kappa\, \widehat{S}^3$ does    not
appear  in  the  renormalizable superpotential of   the  former.  This
particularly simple  form of  the renormalizable MNSSM  superpotential
may     be enforced by     discrete  $R$-symmetries,  such as   ${\cal
  Z}^R_5$~\cite{PT,PP} and ${\cal Z}^R_7$~\cite{PP}.  These   discrete
$R$-symmetries, however,   must  be  extended to  the  gravity-induced
non-renormalizable superpotential   and K\"ahler potential   terms  as
well.    Here,  we   consider  the scenario   of    $N=1$ supergravity
spontaneously  broken   by   a set  of   hidden-sector   fields at  an
intermediate scale.  Within  this framework of  SUSY-breaking, we have
been    able  to show~\cite{PP} that   the    above $R$-symmetries are
sufficient to guarantee   the appearance of the potentially  dangerous
tadpole  $t_S\, S$, with   $t_S  \sim (1/16\pi^2)^n M_{\rm  P}M^2_{\rm
  SUSY}$, at loop levels $n$ higher than 5.  As a consequence, we have
$|t_S| \stackrel{<}{{}_\sim}  1$--10~TeV$^3$,  and therefore the gauge
hierarchy  does  not get  destabilized.   Notice that the so-generated
tadpole  $t_S\, S$ together    with the soft SUSY-breaking mass   term
$m^2_S  S^* S$    lead  to   a VEV for       $S$, $\big<  S\big>     =
\frac{1}{\sqrt{2}}  v_S$, of order   $M_{\rm SUSY}$.  The latter gives
rise to  a $\mu$-parameter  at  the required  electroweak scale, i.e.\ 
$\mu  =  -\frac{1}{\sqrt{2}}\, \lambda  v_S  \sim  M_{\rm SUSY}$, thus
offering a natural explanation for the origin  of the $\mu$-parameter. 
Finally, since the effective  tadpole term $t_S\, S$ explicitly breaks
the   continuous  Peccei--Quinn  symmetry   governing  the   remaining
renormalizable Lagrangian of the  MNSSM,  the theory naturally  avoids
the presence of a phenomenologically excluded weak-scale axion.

The MNSSM predicts,  in addition  to the  charged Higgs scalar  $H^+$,
five neutral Higgs  bosons.   Under the  assumption  of CP invariance,
three of   the neutral Higgs  particles, denoted  as $H_1$,  $H_2$ and
$H_3$  in order of increasing mass,  are CP-even, while the other two,
$A_1$ and $A_2$ (with $M_{A_1} < M_{A_2}$), are CP-odd.  Nevertheless,
since the  tadpole $|t_S|$  naturally  takes values  of  the order  of
1--10~TeV$^3$,  the Higgs-boson mass spectrum  of the MNSSM simplifies
considerably: the  heaviest  states $H_3$ and  $A_2$,  with $M^2_{H_3}
\approx M^2_{A_2} \approx \lambda  t_S/\mu$, decouple as singlets from
the remaining  Higgs sector.  Then,   the masses  of $H^+$ and   $A_1$
satisfy the relation \begin{equation}
  \label{MA1}
 M^2_{A_1}\ \approx\ M^2_a\ =\ M^2_{H^+}\: -\: M^2_W\: +\:
{\textstyle \frac{1}2}\, \lambda^2 v^2\: -\: \delta_{\rm rem}\,,
\end{equation}
where  $M_W = g_w  v/2$ is the  $W$-boson  mass and $\delta_{\rm rem}$
contains   the  radiative corrections   which   may  be  approximately
determined by~\cite{HH,PP,PW}
\begin{equation}
  \label{drem} \delta_{\rm  rem}\  \approx\ -\,\frac{3h^4_t}{32\pi^2}\
\frac{\mu^2v^2}{m^2_{\tilde{t}_1}      +   m^2_{\tilde{t}_2}}\       ,
\end{equation} where $\tilde{t}_1$ and $\tilde{t}_2$ are the stop mass
eigenstates.   Notice that the  relation (\ref{MA1}) is very analogous
to the one known  from the MSSM.   Specifically, the squared mass term
$M^2_a$  enters the non-decoupled $2\times  2$ CP-even mass matrix the
same way as the  squared mass of the  would-be CP-odd Higgs  scalar in
the MSSM.  As opposed  to the MSSM, however,  the presence of the term
$\frac{1}2 \lambda^2 v^2$ in  Eq.~(\ref{MA1})  implies that the  $H^+$
boson can become  even  lighter than $A_1$,   for $\lambda \sim  g_w$;
$H^+$ can  be as light as its  experimental lower bound, $M_{H^+} \sim
80$~GeV~\cite{PDG,ADLO}.  As an important consequence, the $H^+$ boson
can naturally be lighter than the SM-like Higgs boson $H$.  As we will
see, this prediction is very unique for the  MNSSM.  In the MSSM, such
a result may be achieved for unconventionally large values of $|\mu|$,
in  which  case  $\delta_{\rm   rem}$ in Eq.~(\ref{drem})   will start
playing a very analogous r\^ole  as the term $\frac{1}2 \lambda^2 v^2$
in Eq.~(\ref{MA1}) does for the MNSSM.

For  our phenomenological discussion,   we  denote  with  $g_{H_iZZ}$,
$g_{H_iWW}$ and $g_{H_iA_jZ}$ the  strength of the effective $H_iWW$-,
$H_iZZ$- and $H_iA_j Z$-  couplings, respectively normalized to the SM
values  of the $HWW$-,  $HZZ$- and $HZG^0$-  couplings, where $G^0$ is
the would-be Goldstone  boson of  $Z$.  These SM-normalized  effective
couplings obey the unitarity relations: $\sum_{i=1}^3 g^2_{H_iVV} = 1$
and $\sum_{i=1}^3   \sum_{j=1}^2 g^2_{H_iA_jZ}  =  1$,  with  $V=W,Z$.
Moreover,   as a   consequence of a    large   $|t_S|$, the  effective
Higgs-to-gauge-boson couplings satisfy the approximate equalities
\begin{equation}
  \label{compl}
g^2_{H_1VV}\ \approx\ g^2_{H_2A_1Z}\,,\qquad g^2_{H_2VV}\ \approx\
g^2_{H_1A_1Z}\, ,
\end{equation}
which are  essentially identical  to the corresponding complementarity
equalities of   the   MSSM.  We should  remark    that   the relations
(\ref{compl}) are not valid  in the NMSSM, since  the states $H_3$ and
$A_2$ do not  decouple as singlets  from the lightest  Higgs sector in
the latter model.

Our  study of the  MNSSM Higgs-boson mass   spectrum in the decoupling
limit   of  a  large  $|t_S|$   utilizes   renormalization-group  (RG)
techniques developed in~\cite{CQW,CHHHWW,KYS} for the MSSM case and so
improves in several respects  earlier considerations in the NMSSM,  in
which an analogous decoupling  limit   is lacking.  Specifically,   in
addition to the one-loop  stop ($\tilde{t}$) and sbottom ($\tilde{b}$)
corrections, our RG improvement consists in including two-loop leading
logarithms induced  by  QCD and   top- ($t$)  and  bottom- ($b$) quark
Yukawa interactions.  Further,  we    take into account  the   leading
logarithms   originating    from  gaugino   and    higgsino   one-loop
graphs~\cite{HH}, as well   as   we implement  the   potentially large
two-loop contributions  induced  by  the  one-loop   $\tilde{t}$-  and
$\tilde{b}$-  squark  thresholds in  the  $t$- and  $b$-  quark Yukawa
couplings~\cite{CHHHWW}.

In  the  MNSSM and     NMSSM, the SM-normalized   effective  couplings
$g_{H_iVV}$  and the CP-even Higgs-boson   masses $M_{H_i}$ satisfy an
important sum rule:
\begin{eqnarray}
  \label{sumrule}
\sum_{i=1}^3 g^2_{H_iVV}\, M^2_{H_i} \!\!&=&\!\! \Big( M^2_Z
\cos^2 2\beta\, +\, {\textstyle \frac{1}2} 
\lambda^2 v^2 \sin^2 2\beta \Big)\, \bigg( 1\: -\:
\frac{3h^2_t}{8\pi^2}\, t\bigg)\nonumber\\
&&\hspace{-2cm}+\, \frac{3 h^4_t v^2 \sin^4\beta}{8\pi^2}\,
\bigg\{ \bigg(\,1\: +\: \frac{4\alpha_s}{3\pi}\, \frac{X_t}{M_{\rm
    SUSY}}\,\bigg)\, \bigg[\, t\: +\: \frac{X^2_t}{M^2_{\rm   SUSY}}\,
\bigg( 1  - \frac{X^2_t}{12  M^2_{\rm SUSY}}\bigg)\, \bigg]\nonumber\\
&&\hspace{-2cm}  +\, \frac{1}{16\pi^2}\, \Big(   {\textstyle\frac{3}2}
h^2_t\: -\: 32\pi   \alpha_s \Big)\,  \bigg[\,  \frac{2X^2_t}{M^2_{\rm
SUSY}}\,  \bigg( 1 -  \frac{X^2_t}{12 M^2_{\rm  SUSY}}\bigg)  t \: +\:
t^2\,   \bigg]\,   \bigg\}\   +\ {\cal   O}\bigg(\frac{X^6_t}{M^6_{\rm
SUSY}}\bigg)\,,\quad   
\end{eqnarray}  
where  $t= \ln (M^2_{\rm  SUSY}/m^2_t)$, and the strong fine structure
constant $\alpha_s$, the  $t$-quark Yukawa coupling  $h_t$ and the  SM
VEV   $v$ are   to  be evaluated   at   $m_t$.  The mass-coupling  sum
rule~(\ref{sumrule}) is    independent  of  $M_{H^+}$ and  makes   the
definite prediction that the mass of the neutral  Higgs boson $H$ with
SM   coupling to the  $Z$  boson, $g^2_{HZZ}\approx  1$, is completely
specified by  a model-dependent   value  determined from the   RHS  of
Eq.~(\ref{sumrule}). It can thus be estimated from Eq.~(\ref{sumrule})
that in the MNSSM and NMSSM, the  SM-like Higgs-boson mass can reach a
maximum  of $\sim 142$~GeV,  for $\tan\beta  =  2$, $\lambda =  0.65$,
$M_{\rm SUSY} \approx 1$~TeV and $X_t \approx 2.45$~TeV.  In addition,
one should observe that the mass-coupling  sum rule holds true for the
MSSM case as  well, after setting  $\lambda = 0$.  The analytic result
of the RHS  of Eq.~(\ref{sumrule}) is then  in agreement  with the one
computed in~\cite{CHHHWW}, after one follows the suggested RG approach
to  implementing stop  threshold   effects  on  the   $t$-quark Yukawa
coupling.

\begin{table}[t]
\begin{center}
\begin{tabular}{lcccc}
\hline\hline
$M_{H^+}$~[GeV] & ~~~$\tan\beta$~~~ & ~~~$\lambda$~~~ & ~~$\mu$~[GeV]~~ & 
                                               $M_{A_1}$~[GeV]\\
\hline
70  & 1.74  & 0.614  &  $-574$ & 101.8 \\
80  & 2.82  & 0.542  &  $-331$ & 95.7  \\
90  & 4.93  & 0.488  &  $-224$ & 95.4  \\
100 & 7.70  & 0.424  &  $-229$ & 95.9  \\
110 & 10.86 & 0.338  &  $-290$ & 96.2  \\
120 & 14.34 & 0.205  &  $-410$ & 96.4  \\
125.2 & 16.35 & 0.   &  $-825$ & 96.5  \\
\hline\hline
\end{tabular}
\end{center}
\caption{Predictions of the MNSSM~\cite{code}, 
using as inputs: $X_t = 0$, $A_t = A_b$, $M_{\rm
SUSY}= 1$~TeV, $m_{\widetilde{B}} = m_{\widetilde{W}} = 0.3$~TeV,
$m_{\tilde{g}} = 1$~TeV, $\lambda t_S/\mu = 5$~TeV$^2$, $M_{H_1}
\approx 95$~GeV, $M_{H_2}\approx 115$~GeV, $g^2_{H_1ZZ}\approx 0.1$
and $g^2_{H_2ZZ}\approx 0.9$.}\label{tab1}
\end{table}

\begin{figure}[p]
   \leavevmode
 \begin{center}
   \epsfxsize=16.5cm
    \epsffile[0 0 539 652]{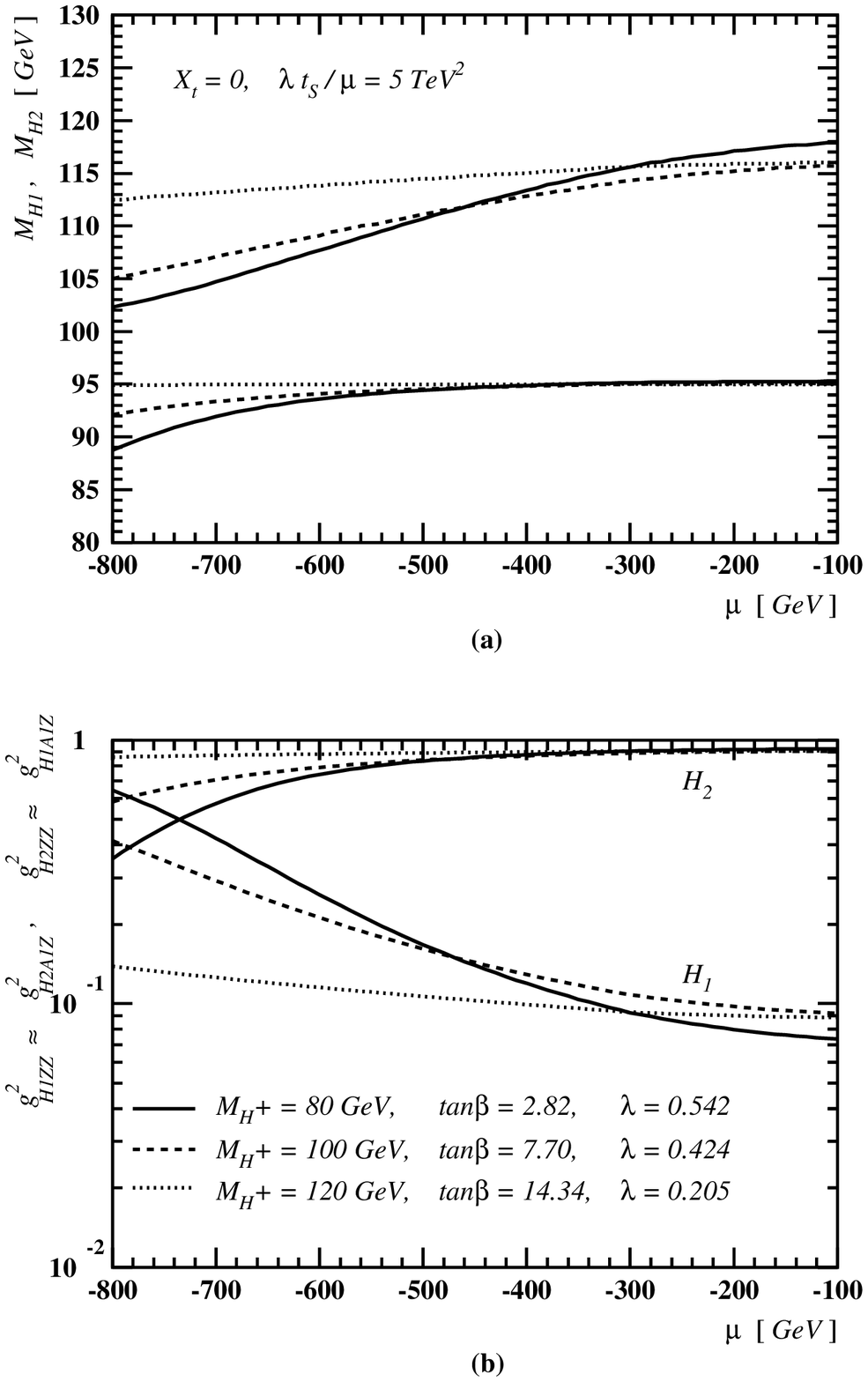}
 \end{center}
 \vspace{-1.2cm} 
\caption{Numerical predictions for (a) $M_{H_1}$ and $M_{H_2}$,
  and (b) $g^2_{H_1ZZ}$ and  $g^2_{H_2ZZ}$,  as functions of $\mu$  in
  the MNSSM.}\label{f1}
\end{figure}

It is now very interesting  to quote results of  variants of the MNSSM
that could  be probed at LEP2  and especially at the upgraded Tevatron
collider in the immediate future.   For definiteness, in our numerical
estimates,  we  fix the soft  SUSY-breaking  squark masses  to $M_{\rm
  SUSY}  = 1$~TeV, and the   U(1)$_Y$, SU(2)$_L$ and SU(3)$_c$ gaugino
masses to  $m_{\widetilde{B}}  =  m_{\widetilde{W}}  =  0.3$~TeV   and
$m_{\tilde{g}} = 1$~TeV,   respectively.   Motivated by the   recently
observed excess of  events for a SM-like  Higgs boson of  a mass $\sim
115$~GeV   at    LEP2~\cite{LEP2},   we choose  in   Tables~\ref{tab1}
and~\ref{tab2}  the mass of  the  second lightest CP-even Higgs  boson
$H_2$ to be $M_{H_2} \approx  115$~GeV, with $g^2_{H_2ZZ}\approx 0.9$. 
For  the lightest Higgs boson $H_1$,  whose squared effective coupling
to the $Z$  boson is necessarily $g^2_{H_1ZZ}  \approx 1 - g^2_{H_2ZZ}
\approx 0.1$, we assume a lower mass,  i.e.\ $M_{H_1} \approx 95$~GeV,
compatible      with    the present    LEP2    data~\cite{ADLO}.    In
Table~\ref{tab1} we consider the zero stop-mixing scenario, i.e.\ $X_t
= 0$, and choose $\lambda t_S/\mu = 5$~TeV$^2$.  We find that the mass
of the  charged  Higgs boson  may   naturally lie below  110~GeV,  for
reasonable values of the  MNSSM parameters.  In particular,  we obtain
$\lambda      \stackrel{<}{{}_\sim}      0.65$,     for     $\tan\beta
\stackrel{>}{{}_\sim} 2$.   Interestingly   enough, such  a   range of
$\lambda$   values is  also  consistent     with the  requirement   of
perturbativity of the MNSSM up to the gauge-coupling unification scale
$M_{\rm U} \sim 10^{16}$~GeV~\cite{CPW,EKW}.  Also, in accordance with
our  earlier discussion, we observe  that the $H^+$  boson  must be as
heavy as 125~GeV in the MSSM limit $\lambda \to 0$, i.e.\ heavier than
the SM-like  Higgs  boson  $H_2$.    In Table~\ref{tab2},  we   select
$\lambda t_S/\mu =  1.5$~TeV$^2$  and use the   value of maximal  stop
mixing,  $X_t \approx \sqrt{6}\,M_{\rm  SUSY}$,  characterized by  the
fact  that  the radiative  effects  given  in Eq.~(\ref{sumrule})  get
approximately  maximized.   We arrive   at the  very  same conclusion,
namely the $H^+$   boson can be  lighter than  $\sim   110$~GeV and so
lighter  than the SM-like Higgs  boson $H_2$.  This particular feature
of   the MNSSM  is   also reflected in   Fig.~\ref{f1},  where we show
numerical   values  of $H_1$-  and $H_2$-  boson   masses and of their
squared effective couplings  to the $Z$ boson  as  functions of $\mu$,
for three variants of the MNSSM from Table~\ref{tab1}: $M_{H^+} = 80$,
100 and  120~GeV.   We observe that  the aforementioned LEP2-motivated
scenario of a SM-like Higgs boson may be accounted for by a wide range
of $\mu$ values.

\begin{table}[t]
\begin{center}
\begin{tabular}{lcccc}
\hline\hline
$M_{H^+}$~[GeV] & ~~~$\tan\beta$~~~ & ~~~$\lambda$~~~ & ~~$\mu$~[GeV]~~ & 
                                               $M_{A_1}$~[GeV]\\
\hline
60  & 2.24   & 0.649 &  $-441$ & 102.8 \\
70  & 3.28   & 0.590 &  $-402$ & 97.8  \\
80  & 5.15   & 0.541 &  $-397$ & 96.7  \\
90  & 7.73   & 0.489 &  $-424$ & 96.7  \\
100 & 10.7   & 0.427 &  $-481$ & 97.1  \\
110 & 14.0   & 0.348 &  $-585$ & 97.7  \\
120 & 17.65  & 0.249 &  $-810$ & 99.0  \\
\hline\hline
\end{tabular}
\end{center}
\caption{Predictions of the  MNSSM~\cite{code},  
using   the  same inputs  as   in
Table~\ref{tab1}, with the exception that $X_t = 2.45$~TeV and
$\lambda t_S/\mu = 1.5$~TeV$^2$.}\label{tab2}
\end{table}

\begin{figure}[p]
   \leavevmode
 \begin{center}
   \epsfxsize=16.5cm
    \epsffile[0 0 539 652]{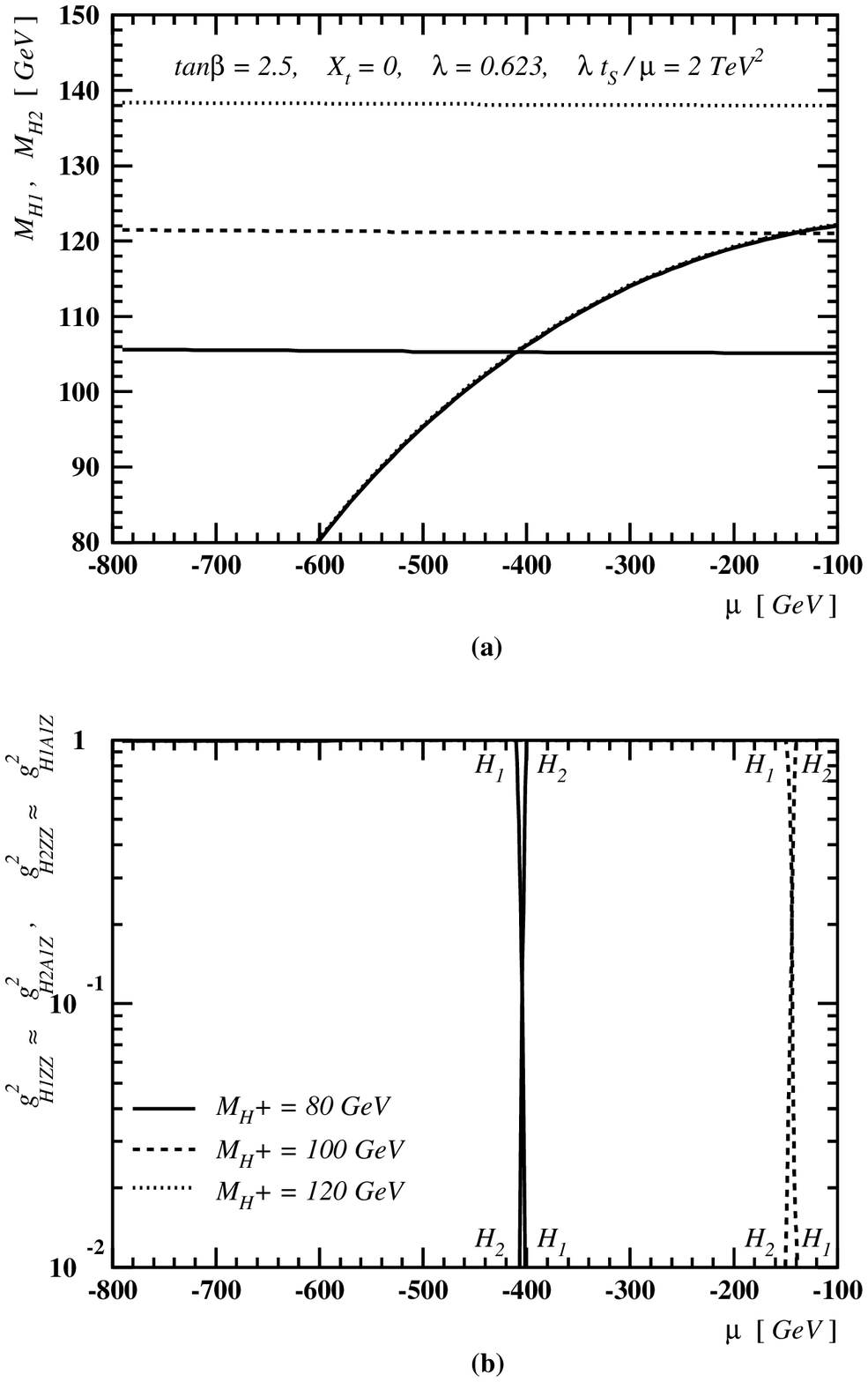}
 \end{center}
 \vspace{-1.2cm} 
\caption{Numerical values of (a) $M_{H_1}$ and $M_{H_2}$,
  and (b) $g^2_{H_1ZZ}$ and  $g^2_{H_2ZZ}$,  as functions of $\mu$  in
  the MNSSM.}\label{f2}
\end{figure}

An  interesting  alternative emerges  if one of  the two non-decoupled
CP-even Higgs bosons, e.g.\ $H_2$, has a mass $M_{H_2}\approx 115$~GeV
with $g^2_{H_2ZZ} \approx  1$, while  the other  one, $H_1$, does  not
couple to the  $Z$ boson but  has $g^2_{A_1H_1Z} \approx  1$, and vice
versa. Such an alternative is easily compatible with the LEP2 data, as
long as   the  mass    inequality  constraint,  $M_{H_1}   +   M_{A_1}
\stackrel{>}{{}_\sim} 170$~GeV~\cite{ADLO}, is met.  Assuming $\lambda
t_S/\mu  = 2$~TeV$^2$, the above scenario  may  be realized for a wide
range of  charged Higgs-boson masses between 60  and  110~GeV, and for
both zero and maximal stop mixing.   For instance, for $X_t=0$, such a
kinematic  dependence  insensitive to $M_{H^+}$  may  be  obtained for
$\tan\beta = 2.5$, $\lambda = 0.623$ and $\mu \approx -283$~GeV, while
for $X_t =  2.45$~TeV, one  may choose  $\tan\beta  = 5$, $\lambda   =
0.645$ and $\mu \approx  -393$~GeV.  In Fig.~\ref{f2}, we display  the
dependence of the $H_1$- and $H_2$- boson  masses and of their squared
effective  couplings  to   the $Z$    boson  as  a   function of   the
$\mu$-parameter,  for  $M_{H^+}=80,\    100$   and  $120$~GeV  in  the
aforementioned variant of the MNSSM   with $X_t=0$.  We observe  again
that the $H^+$ boson can be  lighter than $\sim 110$~GeV and therefore
lighter than  the  SM-like Higgs boson.  In   addition, we notice that
there exists a   SM-like Higgs boson for a   very wide range of  $\mu$
values and, only for  a very short  interval  of $\mu$, the  $H_1$ and
$H_2$ bosons interchange their couplings to  the $Z$ boson, while they
are degenerate  in mass.  Finally,  we should reiterate the  fact that
analogous possibilities  are present in  the MSSM  for large values of
$|\mu|$.  In agreement with our earlier observation,  we find that the
$H^+$ boson   can  be  as  light  as  100~GeV,  with  $M_{H_2} \approx
115$~GeV,  $M_{H_1} \approx  82.3$~GeV,  $M_{A_1}\approx 92.6$~GeV and
$g^2_{H_2ZZ} \approx g^2_{H_1A_1Z}  \approx 1$, if  the large value of
$|\mu|$, $\mu \approx -3.97$~TeV,  together with $\tan\beta=12.3$  and
$X_t=1$~TeV, is employed.

In the NMSSM the situation is different.  The charged Higgs-boson mass
$M_{H^+}$   exhibits    a    strong  monotonic   dependence    on  the
$\mu$-parameter; $|\mu|$ gets rapidly   smaller for smaller values  of
$M_{H^+}$.  This generic feature of the NMSSM may mainly be attributed
to   the fact~\cite{PP} that  no analogous  decoupling limit  due to a
large $|t_S|$ exists in  this model.  In  particular, unlike the MNSSM
case, no actual use  of the  presence  of the  contribution $\frac{1}2
\lambda^2 v^2$  in the Higgs-boson mass  matrices  can be made  in the
NMSSM.  In      fact,  we    find  that     it  is  always    $M_{H^+}
\stackrel{>}{{}_\sim} 110$~GeV, for phenomenologically favoured values
of     $\mu$,        i.e.~for         $|\mu|     \stackrel{>}{{}_\sim}
100$~GeV~\cite{PDG,GKLR},\footnote[2]{If this   last constraint on the
  $\mu$-parameter is lifted, then charged Higgs-boson masses as low as
  90~GeV might be  possible in the NMSSM~\cite{Drees}.}  assuming that
the  theory stays perturbative up to  $M_{\rm U}$.  If the $H^+$ boson
becomes  heavier  than  the  neutral  SM-like   Higgs boson  $H$,  the
phenomenological distinction between the   NMSSM and MNSSM  is getting
more    difficult  and additional experimental    information would be
necessary,    such as  the  testing  of  the  complementarity coupling
relations of Eq.~(\ref{compl}).

To summarize: the renormalizable low-energy sector of the MNSSM in the
decoupling limit due to a  large $|t_S|$ has effectively one parameter
more  than the corresponding one of   the (CP-conserving) MSSM, namely
the coupling $\lambda$.  In fact, in the MNSSM the natural size of the
higher-loop  generated   tadpole   parameter   $|t_S|$  is   of  order
1--10~TeV$^3$.  For unsuppressed  values of $\lambda$, $t_S$ leads  to
masses of the order of 1 TeV for the heaviest CP-even and CP-odd Higgs
scalars $H_3$ and $A_2$,  so these states  decouple as heavy  singlets
giving rise  to an active  low-energy Higgs sector  consisting only of
doublet-Higgs fields,  closely analogous to the one  of the MSSM.  The
latter should be  contrasted with the  NMSSM case, where no  analogous
decoupling  limit due to  a large $|t_S|$  exists in this model.  Most
strikingly,  the MNSSM may also  predict a light  charged Higgs boson,
which can be even lighter than the SM-like Higgs boson $H$.  We should
stress  again that in   the light of the   present  LEP2 data, such  a
prediction cannot be naturally obtained in the MSSM  or NMSSM.  In the
same vein, we note that it would be very interesting  to study as well
as  identify the     compelling low-energy structure  of   other  SUSY
extensions of the  SM that could   lead to the inverse mass  hierarchy
$M_{H^+} \ll M_H$.  {}From our discussion, however, it is obvious that
the MNSSM trully represents the simplest and most economic non-minimal
supersymmetric model proposed in  the literature after  the MSSM.   In
conclusion,  it is very important that  the upgraded Tevatron collider
has the physics potential to probe the  exciting hypothesis of a light
charged Higgs boson in top decays $t\to H^+b$~\cite{Hplus} and analyze
its possible consequences within the framework of the MNSSM.

\subsection*{Acknowledgements}
The work of A.P.\ is supported in part by the Bundesministerium f\"ur
Bildung und Forschung under the contract number 05HT9WWA9.

\newpage

\begin{thebibliography}{99}
  
\bibitem{SUSY} For reviews, see, H.-P. Nilles, Phys.\ Rep.\ {\bf 110}
  (1984) 1; H.E. Haber and G.L. Kane, Phys.\ Rep.\ {\bf 117} (1985)
  75; A.B. Lahanas and D.V. Nanopoulos, Phys.\ Rep.\ {\bf 145}
  (1987) 1; J.F. Gunion, H.E. Haber, G.L. Kane and S. Dawson, ``The Higgs
  Hunter's Guide,'' (Addison-Wesley, Reading, MA, 1990).

\bibitem{CPW} See, e.g., M. Carena, S. Pokorski and C.E.M. Wagner,
  Nucl.\ Phys.\ {\bf B406} (1993) 59.
  
\bibitem{PDG} Review of Particle Physics (D.E. Groom et al.), Eur.\ 
  Phys.\ J. {\bf C15} (2000) 1.
  
\bibitem{EQ} For example, see, J.R. Espinosa and M. Quir\'os, Phys.\ 
  Rev.\ Lett.\ {\bf 81} (1998) 516.
  
\bibitem{SHiggs} J.  Ellis, G.  Ridolfi and F.  Zwirner, Phys.\ Lett.\ 
  {\bf B257} (1991) 83; M.S.  Berger, Phys.\ Rev.\ {\bf D41} (1990)
  225; Y.  Okada, M.  Yamaguchi and T. Yanagida, Prog.\ Theor.\ Phys.\ 
  {\bf 85} (1991) 1; H.E.  Haber and R.  Hempfling, Phys.\ Rev.\ 
  Lett.\ {\bf 66} (1991) 1815.
  
\bibitem{CQW} M. Carena, M.  Quir\'os and C.E.M. Wagner, Nucl.\ Phys.\
  {\bf B461} (1996) 407; M.  Carena, S.  Mrenna and C.E.M.  Wagner,
  Phys.\ Rev.\ {\bf D60} (1999) 075010.
  
\bibitem{HHW} S. Heinemeyer, W. Hollik and G.  Weiglein, Phys.\ Rev.\
{\bf D58} (1998) 091701.

\bibitem{CHHHWW} M.  Carena, H.E.  Haber, S.  Heinemeyer, W.   Hollik,
  C.E.M.  Wagner  and G.  Weiglein,  Nucl.\ Phys.\ {\bf B580} (2000) 29;
  J.R.  Espinosa and R.-J.  Zhang, JHEP {\bf 0003} (2000) 026.
  
\bibitem{nmssm} P.  Fayet, Nucl.\ Phys.\ {\bf B90} (1975) 104; H.-P.
  Nilles, M.  Srednicki and D.  Wyler, Phys.\ Lett.\ {\bf B120} (1983)
  346; A.B. Lahanas, Phys.\ Lett.\ {\bf B124} (1983) 341. For reviews,
  see, M. Drees, Int.\ J. Mod.\ Phys.\ {\bf A4} (1989) 3635; F. Franke
  and H.  Fraas, Int.\ J. Mod.\ Phys.\ {\bf A12} (1997) 479.
  
\bibitem{EKW} T.  Elliott, S.F. King and P.L.  White, Phys.\ Rev.\ 
  {\bf D49} (1994) 2435. For a recent computation, see, U. Ellwanger
  and C. Hugonie, hep-ph/9909260.
  
\bibitem{ADLO} ALEPH, DELPHI, L3 and OPAL Collaborations (P.  Bock et
  al.), CERN-EP-2000-055; OPAL Collaboration (G. Abbiendi et al.),
  CERN-EP-2000-092, hep-ex/0007040.

\bibitem{PP} C. Panagiotakopoulos and A. Pilaftsis, Phys.\ Rev.\ {\bf
    D63} (2001) 055003.

\bibitem{GKLR} N. Ghodbane, S. Katsanevas, I. Laktineh and J. Rosiek,
  hep-ph/0012031.
  
\bibitem{BBMR} S. Bertolini, F. Borzumati, A. Masiero and G. Ridolfi,
  Nucl.\ Phys.\ {\bf B353} (1991) 591; R. Barbieri and G.F. Giudice,
  Phys.\ Lett.\ {\bf B309} (1993) 86; R. Garisto and J.N. Ng, Phys.\ 
  Lett.\ {\bf B315} (1993) 372.
  
\bibitem{NS} Y. Nir and N.  Seiberg, Phys.\ Lett.\ {\bf B309} (1993)
  337; D.  Choudhury, F.  Eberlein, A.~K\"onig, J.  Louis and S.
  Pokorski, Phys.\ Lett.\ {\bf B342} (1995) 180.

\bibitem{PT} C. Panagiotakopoulos and K. Tamvakis, Phys.\ Lett.\ {\bf
    B469} (1999) 145.
  
\bibitem{HH} H. Haber and R. Hempfling, Phys.\ Rev.\ {\bf D48} (1993)
  4280.
  
\bibitem{PW} For a RG-improved computation of $M_{H^+}$ in the MSSM,
  see, A.  Pilaftsis and C.E.M.~Wagner, Nucl.\ Phys.\ {\bf B553}
  (1999) 3; M.  Carena, J.  Ellis, A.  Pilaftsis and C.E.M.~Wagner,
  Nucl.\ Phys.\ {\bf B586} (2000) 92; Phys.\ Lett.\ {\bf B495} (2000)
  155.
  
\bibitem{KYS} J. Kodaira, Y. Yasui and K. Sasaki, Phys.\ Rev.\ {\bf
    D50} (1994) 7035; J.A. Casas, J.R. Espinosa, M. Quir\'os and A.
  Riotto, Nucl.\ Phys.\ {\bf B436} (1995) 3; Nucl.\ Phys.\ {\bf B439}
  (1995) 466 (Erratum).

\bibitem{LEP2} ALEPH Collaboration (R. Barate et al.), Phys.\ Lett.\ 
  {\bf B495} (2000) 1; L3 Collaboration (M. Acciarri et al.), Phys.\ 
  Lett.\ {\bf B495} (2000) 18.
    
\bibitem{code} Numerical values are obtained by the Fortran code {\tt
 mnssm} which is available from {\tt http://pilaftsi.home.cern.ch/pilaftsi/}.

\bibitem{Drees} M.  Drees, E.  Ma,  P.N.  Pandita, D.P. Roy  and  S.K. 
  Vempati, Phys.\ Lett.\ {\bf B433} (1998) 346.

\bibitem{Hplus} T. Affolder et al., The CDF Collaboration, {\tt
    http://www-cdf.fnal.gov/physics/
    preprints/cdf5124\_charged\_higgs\_prd.ps}.

\end{thebibliography}
\end{document}